\documentclass[11pt,reqno]{amsart}

\newtheorem{theorem}{Theorem}

\usepackage{pst-all,amsthm,graphicx,amsmath,amssymb}
\usepackage[latin1]{inputenc}
\usepackage{mathrsfs}
\usepackage{dsfont}
\usepackage{hyperref}

\newcommand{\cL}{\mathcal{L}}

\newcommand\C{{\mathbb C}}
\newcommand\R{{\mathbb R}}

\renewcommand\Tr{{\rm Tr}}
\newcommand\half{{\mbox{$\frac 12$}}}

\newcommand{\beq}{\begin{equation}}
\newcommand{\eeq}{\end{equation}}

\newcommand\infspec{{\rm inf\, spec\, }}
\def\x{{\vec{x}}}
\def\y{{\vec{y}}}

\renewcommand\rho\varrho
\newcommand\str[1]{{\bf #1}}

\begin{document}

\title{HOT TOPICS IN COLD GASES}

\thanks{Plenary lecture given at the XVI International Congress on Mathematical Physics, Prague, August 3--8, 2009.\\ \copyright\, 2009 by the author. This article may be reproduced, in its entirety, for non-commercial purposes.}

\author{R. Seiringer}

\address{R. Seiringer, Department of Physics, Princeton University, Princeton NJ 08542, USA. 
E-mail: rseiring@princeton.edu}

\begin{abstract}
  Since the first experimental realization of Bose-Einstein
  condensation in cold atomic gases in 1995 there has been a surge of
  activity in this field. Ingenious experiments have allowed us to
  probe matter close to zero temperature and reveal some of the
  fascinating effects quantum mechanics has bestowed on nature. It is
  a challenge for mathematical physicists to understand these various
  phenomena from first principles, that is, starting from the
  underlying many-body Schr\"odinger equation. Recent progress in this
  direction concerns mainly equilibrium properties of dilute, cold
  quantum gases. We shall explain some of the results in this article,
  and describe the mathematics involved in understanding these
  phenomena. Topics include the ground state energy and the free
  energy at positive temperature, the effect of interparticle
  interaction on the critical temperature for Bose-Einstein
  condensation, as well as the occurrence of superfluidity and
  quantized vortices in rapidly rotating gases.
\end{abstract}

%\keywords{Bose-Einstein condensation; statistical mechanics; dilute gases; superfluidity.}

\maketitle

\section{Introduction}
Bose-Einstein Condensation (BEC) was first experimentally realized in
cold atomic gases in 1995 \cite{AEMWC,DMAVDKW}. In these experiments,
a large number of (bosonic) atoms is confined to a trap and cooled to
very low temperatures. Below a certain critical temperature
condensation of a large fraction of particles into the same
one-particle state occurs.

These Bose-Einstein condensates display various interesting quantum
phenomena, like superfluidity and the appearance of quantized
vortices in rotating traps, effective lower dimensional behavior in
strongly elongated traps, etc. We refer to the review articles
\cite{DGPS,Zwerger,Cooper,Fetter} for an overview of the
state-of-the-art of this subject and a list of references to the
original literature.

BEC was predicted by Einstein in 1924 \cite{Einstein} from
considerations of the non-interacting Bose gas, extending the
work of Bose \cite{Bose} to massive particles. The presence of
particle interactions represents a major difficulty for a rigorous
derivation of this phenomenon, however, as we shall discuss below.

\subsection{The Bose Gas: A Quantum Many-Body Problem}

The quantum-mechanical description of the Bose gas is given in terms
of its Hamiltonian.  For a gas of $N$ bosons confined to a region
$\Lambda\in \R^3$, and interacting via a repulsive pair-interaction
potential $v$, it is given by
\begin{equation}\label{ham}
\boxed{\ 
H = - \sum_{i=1}^{N}  \Delta_i + \sum_{1 \leq i < j \leq N} v(\x_i - \x_j) \ }
\end{equation}
The kinetic energy of the particles is described by the Laplacian
$\Delta=\vec\nabla^2$, and we choose Dirichlet boundary conditions on
$\partial\Lambda$ for concreteness. Other boundary conditions could be
used as well. The subscript $i$ stands for the action in the $i$th
particle coordinate $\x_i\in \R^3$. Units are chosen such that $\hbar
= 2m=1$, with $m$ the mass of the particles.

The Hamiltonian $H$ acts on the Hilbert space of permutation-symmetric
wave functions $\Psi\in \bigotimes^N L^2(\R^3)$, as appropriate for
bosons; i.e., square-integrable functions of $N$ variables $\x_i\in
\R^3$ satisfying
$$
\psi(\x_1,\dots,\x_N) = \psi(\x_{\pi(1)},\dots,\x_{\pi(N)}) 
$$
for any permutation $\pi$ of $(1,2,\dots, N)$. 

In the following, the interaction $v$ will be assumed to be radial and
non-negative. Moreover, it is sufficiently short range as to have a
finite scattering length, which means that it is integrable outside
some compact set. No other regularity assumptions will be made. In
particular, $v$ is allowed to have a hard core, which reduces the
domain of definition of $H$ to those functions $\Psi$ that vanish
whenever the distance between a pair of particle coordinates is
smaller than the hard-sphere radius. A particular example of an
interaction potential to keep in mind are pure hard spheres where,
formally, $v(\x)=\infty$ for $|\x|\leq a$, and $v(\x)=0$ for $|\x|>a$.

The present setup can be easily generalized to describe inhomogeneous
systems in a trap. One simply adds trap potential 
$$\sum_{i=1}^N V(\x_i)$$ 
to $H$, where $V$ is a real-valued, locally bounded function
with $\lim_{|\x|\to\infty} V(\x)=\infty$. The latter condition
guarantees that the particles are confined to the trap, even in case
$\Lambda=\R^3$.

Similarly, rotating systems can be described adding the term
$$\sum_{i=1}^N \vec\Omega\cdot \vec L_i$$ 
to the Hamiltonian $H$, with $\vec\Omega\in \R^3$ being the angular
velocity and $\vec L = -i \x \wedge\vec \nabla $ the angular momentum
operator. This term results from a transformation to the rotating
frame of reference.

\subsection{Quantities of Interest}

In the following, we shall distinguish two types of questions that can
be asked concerning of the behavior of Bose gases described by the
Hamiltonian (\ref{ham}) above.

\begin{itemize}

\item \str{Thermodynamic quantities}, like the ground state energy per
  unit volume, or the free energy density at positive
  temperature. Here one considers homogeneous systems and is
  interested in the thermodynamic limit $N\to \infty$, $\Lambda\to
  \R^3$ with the particle density $\rho= N/|\Lambda|$ fixed. 

  Of particular interest is the notion of Bose-Einstein condensation,
  which concerns off-diagonal long-range order in the one-particle
  density matrix $\langle a^\dagger(x) a(y) \rangle$, and is expected
  to occur below a critical temperature.

\item \str{Behavior of trapped systems} in the ground state. One
  observes interesting quantum phenomena, like effective
  one-dimensional behavior in strongly elongated traps, vortices in
  rotating systems, a bosonic analogue of the fractional quantum Hall
  effect in rapidly rotating gases, etc.

  Of particular relevance is the \str{Gross-Pitaevskii} scaling, where
  the ratio of the scattering length $a$ to the diameter of the trap
  is $O(N^{-1})$.

\end{itemize}

We shall discuss our current knowledge about answers to these
questions, as far as mathematical physics is concerned, in the
following sections.

\section{Homogeneous Systems in the Thermodynamic Limit}

\subsection{The Ground State Energy of Homogeneous Bose Gases}
Consider first the case of a homogeneous system in the absence of a
trapping potential or rotation. The ground state energy density in the
thermodynamic limit is given by
\begin{equation}
e(\rho) = \lim_{\Lambda\to \R^3, \, N/|\Lambda|\to \rho} \frac
1{|\Lambda|}\inf{\rm \,spec\,}H
\end{equation}
with $H$ as in (\ref{ham}). The existence of this thermodynamic limit
is well understood for appropriate sequences of domains $\Lambda$
approaching $\R^3$. See, e.g., Ruelle's book \cite{Ruelle}.

We will be particularly interested in the limit of low density, when the
gas is dilute in the sense that $a^3\rho\ll 1$, where $a$ denotes the
scattering length of the interaction potential $v$. It is defined as
\begin{equation}\label{defa}
4\pi a =\inf \left\{  \int_{\R^3} \left(
     |\vec\nabla\phi(|\x|)|^2 + \half v(\x) \phi(|\x|)^2\right)d\x \, :
   \, \phi\geq 0\,,\,
   \lim_{r\to\infty} \phi(r) = 1\right\}\,.
\end{equation}
For bosons at low density, one expects that 
\begin{equation}\label{erho}
e(\rho) \approx 4\pi a
\rho^2 \,.
\end{equation} 
This formula is suggested by considering the ground state
energy of two bosons in a large region $\Lambda$, which is $8\pi
a/|\Lambda|$, as can be easily deduced from (\ref{defa}). Multiplying
this by the number of pairs of bosons, $N(N-1)/2$, one arrives at
(\ref{erho}). That this simple heuristics is
correct is far from obvious, however. It fails for two-dimensional
systems, for instance \cite{Schick,LY2001}.

The investigation of the ground state energy density $e(\rho)$ goes
back to Bogoliubov \cite{B1947} in the 40s, and Lee, Huang and Yang in
the 50s \cite{LHY}.  Dyson \cite{D1957} computed a rigorous upper
bound that shows the correct leading order asymptotics (\ref{erho})
for hard spheres, but his lower bound was 14 times too small. His
upper bound was later generalized to arbitrary repulsive interaction
potentials in \cite{LSY00}. The correct lower bound was proved only
in 1998 by Lieb and Yngvason \cite{LY1998}. We formulate this result
as a theorem.

\begin{theorem}[\str{Bosons at $T=0$}] As $\rho \to 0$, 
\begin{equation}\label{dly}
 \boxed{\ e(\rho) = 4\pi a \rho^2 + o(\rho^2) \ }
\end{equation}
\end{theorem}

Note that if one treats the interaction energy as a perturbation of
the kinetic energy, naive perturbation theory would yield $\half \int
v$ instead of $4\pi a$. This is always too big, as (\ref{defa}) shows,
and would even be infinite for hard spheres.  In fact, the result
(\ref{dly}) is non-perturbative in the sense that the scattering
length $a$ contains terms to arbitrary high order in the interaction
potential $v$.

It remains an open problem to establish the leading order correction to (\ref{erho}), which is expected to be given by the Lee-Huang-Yang formula \cite{LHY}
\begin{equation}\label{lhy}
e(\rho) \approx 4\pi a\rho^2\left(1+ \frac{128}{15\sqrt{\pi}} \sqrt{a^3\rho}\right)\,.
\end{equation}
Recent progress in this direction was made in \cite{GiulianiS09} and
\cite{LiebSolovej09}, where it was shown that (\ref{lhy}) holds for
certain density-dependent and appropriately scaled interaction
potentials. The general question remains open, however.

\subsection{Homogeneous Bose Gas at Positive Temperature}

At positive temperature $T>0$, the appropriate quantity to consider is the  free energy density, which is defined as 
\begin{equation}
f(\rho,T) = -  T\lim_{\Lambda\to \R^3, \, N/|\Lambda|\to \rho}
  \frac 1{|\Lambda|} \ln \Tr \exp(-H/T) \,.
\end{equation}
For non-interacting bosons (i.e., $v\equiv 0$), it can be calculated explicitly. We denote it by $f_0(\rho,T)$. It is given in terms of a Legendre transform as  
\begin{equation}\label{deff0}
f_0(\rho,T) = \sup_{\mu <0} \left[ \mu \rho +
  \frac T{(2\pi)^3} \int_{\R^3} d\vec p\,
\ln\left(1-\exp(-({\vec p}^2-\mu)/T)\right)\right] \,.
\end{equation}
Note that it has the scaling property
$$
f_0(\rho,T) = \rho^{5/3}f_0(1,T\rho^{-2/3}) 
$$
which follows from the fact that the only length scales in the problem
are the mean particle spacing $\rho^{-1/3}$ and the thermal wavelength
$T^{-1/2}$, and hence $f_0$ depends, up to a prefactor, only on
their ratio.

From (\ref{deff0}) it is easy to see that $f_0$ is not an analytic function of $\rho$ (or $T$), and hence even a non-interacting Bose gas shows a phase transition. This transition is known as Bose-Einstein condensation, and occurs at a critical density 
$$\rho_c(T) = \zeta(\tfrac 32) \left(\frac T{4\pi} \right)^{3/2}\,,
$$
where $\zeta$ denotes the Riemann zeta-function. 
In fact, $\partial f_0(\rho,T)/\partial\rho = 0$ for $\rho\geq
\rho_c$. For $\rho>\rho_c(T)$, $\rho-\rho_c(T)$ is interpreted as the
density of the Bose-Einstein condensate.

For interacting gases, there are now three length scales to consider:
the interaction range $a$, the mean particle distance $\rho^{-1/3}$,
and the thermal wavelength $T^{-1/2}$. For dilute systems, one
considers the case
$$ a \ll \rho^{-1/3} \sim T^{-1/2} \,.$$ In this regime, the free energy turns out to be the given by the following expression.

\begin{theorem}[\str{Bosons at $T>0$}]\label{thmT}
 For $a^3\rho\ll 1$ we have
\begin{equation}\label{fT}
\boxed{\quad f(\rho,T) = f_0(\rho,T) + 4\pi a  \left( 2 \rho^2
    - \left[ \rho-\rho_c(T)\right]_+^2\right)    
+ o(\rho^2) \quad}
\end{equation}
where $[t]_+=\max\{t,0\}$ denotes the positive part.
\end{theorem}

The lower bound in (\ref{fT}) was proved in \cite{S08}. An upper bound
for smooth interacting potentials of rapid decay was later obtained
in \cite{Yin2009}, the more general case being still open.

The error term in (\ref{fT}) is uniform in $T/\rho^{2/3}$ for bounded $T/\rho^{2/3}$, corresponding to the quantum regime. For $T/\rho^{2/3} \to \infty$ one obtains a {\it classical} gas, whereas for $T/\rho^{2/3} \to 0$ the system approaches the ground state. 

Note that for $\rho<\rho_c(T)$, the leading order correction compared
to the ideal Bose gas is $8\pi a \rho^2$ instead of the $4\pi a
\rho^2$ at zero temperature. The additional factor $2$ is a result of
the symmetry requirements of the wave functions and can be interpreted
as an exchange term; this symmetrization applies only to particles
outside the condensate, however, and this explains the subtraction of
the square of the condensate density in (\ref{fT}). We also remark
that without restricting to symmetric functions, the leading order
correction compared with an ideal gas would be $4\pi a\rho^2$ at any
$T>0$, just like at $T=0$.

The proof of Theorem~\ref{thmT} is long and technical and hence can not be reproduced here. One of the key issues to understand is a certain separation of energy scales in the two terms on the right side of (\ref{fT}). In momentum space, these are

\begin{itemize}

\item large momenta $|\vec p| \sim
1/a$ responsible for scattering of two particles at a distance $\sim a$ of each other.

\item low momenta  $|\vec p|\sim T^{1/2}\ll 1/a$, responsible for the thermal distribution distribution of the particles' kinetic energy.

\item Bose-Einstein condensation at momentum $\vec p = 0$.

\end{itemize}

\subsection{Critical Temperature for BEC}

As discussed above, the ideal, non-interacting Bose gas displays a phase transition above a critical density. 
Equivalently, BEC in the ideal gas occurs below the critical temperature 
$$T_c(\rho) = \frac{4\pi}{\zeta(3/2)^{2/3}} \rho^{2/3}\,.$$

A useful characterization of BEC, applicable also for interacting systems, is in terms of the one-particle density matrix of the system. This density matrix is defined as 
\begin{equation}\label{def:rdm}
\gamma = N \frac 1 {\Tr\, e^{-H/T}} \Tr^{(N-1)} e^{-H/T}
\end{equation}
where $\Tr^{N-1}$ stands for the partial trace over $N-1$ particle
coordinates. Hence $\gamma$ is an operator on the one-particle space
$L^2(\R^3)$. Obviously $\gamma\geq 0$ and $\Tr\, \gamma = N$, by
definition. BEC is characterized by the fact that, in the
thermodynamic limit, the integral kernel $\gamma(\x,\y)$ of $\gamma$
does not vanish as $|\x-\y|\to \infty$. This is also referred to as
off-diagonal long range order. For non-interacting bosons, one can
show that
$$
\gamma(\x,\y) = \left[\rho-\rho_c(T)\right]_+ + \sum_{n\geq 0} \frac{e^{\bar\mu 
n/T}}{(4\pi n/T)^{3/2} } e^{-T |\x-\y|^2/(4 n)} 
$$
in the thermodynamic limit, with $[t]_+=\max\{t,0\}$ denoting the positive part, and $\bar\mu\leq 0$ the $\mu$ where the maximum in (\ref{deff0}) is achieved. Hence the kernel $\gamma(\x,\y)$ has the following characteristics:

\begin{itemize}
\item For $T<T_c(\rho)$, $\gamma(\x,\y)$ does not decay. In fact, $\lim_{|\x-\y|\to\infty} \gamma(\x,\y) = \rho - \rho_c(T)$, the condensate density
\item For $T>T_c(\rho)$, $\gamma(\x,\y)$ decays exponentially, like $e^{-\sqrt{-\bar\mu}|\x-\y|}$
\item For $T=T_c(\rho)$, $\gamma(\x,\y)$ decays algebraically. In fact, $\gamma(\x,\y) \sim |\x-\y|^{-1}$ in this case.
\end{itemize}

These features are expected to hold also for interacting Bose gases,
although with a different value of the critical temperature
$T_c(\rho)$.  It is still an open problem to prove the existence of
BEC for interacting gases, however. The only known case where BEC has
been proved is the hard-core lattice gas at half filling
\cite{DysonLiebSimon1978}, which is equivalent to the XY spin model.\cite{MM}

Although there is no proof that $T_c\neq 0$ in the interacting case,
an upper bound can be derived rigorously \cite{SUeltschi2009}:

\begin{theorem}[\str{Upper bound on $T_c$}]\label{tcthm}
For small $a^3\rho$ and some $c>0$, 
$$
\boxed{ \ \frac{ T_c - T_c^{(0)} } {T_c^{(0)}} \leq c \sqrt{ a \rho^{1/3} } \ }
$$
where $T_c^{(0)} = \frac{4\pi}{\zeta(3/2)^{2/3}} \rho^{2/3}$ is the critical temperature for the ideal Bose gas.
\end{theorem}

More precisely, it is shown in \cite{SUeltschi2009} that
$\gamma(\x,\y)$ decays exponentially if $T> T_c^{(0)}( 1 + c
\sqrt{a\rho^{1/3}})$. The proof uses a well-known Feynman-Kac
representation of the partition function in terms of integrals over
paths and sums over cycles in permutations.\cite{ginibre}

\begin{figure}[ht]
\begin{center}
\includegraphics[height=.5\hsize]{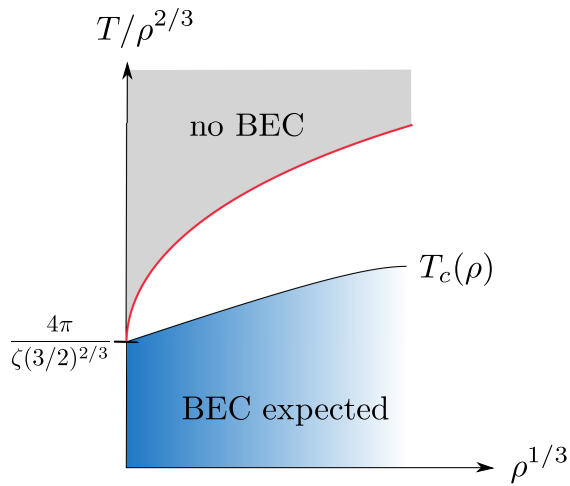}
\end{center}
\caption{The red line shows the rigorous upper bound on the critical temperature for BEC. The dashed line corresponds to the expected behavior based on numerical simulations.}\label{figtc}
\end{figure}

There seems to be still no consensus in the physics literature
concerning the correct power of the exponent of $a\rho^{1/3}$ in the
shift in critical temperature, or even the sign of $c$!  Recent
numerical simulations suggest that the shift should be linear in
$a\rho^{1/3}$, with a positive $c$. This expected behavior of
$T_c(\rho)$, as well as the upper bound of Theorem~\ref{tcthm}, are
sketched in Figure~\ref{figtc}.

\section{Trapped Bose Gases}

In the previous chapter we considered homogeneous Bose gases in the
thermodynamic limit. Recent experiments with cold atoms consider
inhomogeneous gases in traps, however. That is, one can take $\Lambda$
to be the whole of $\R^3$, but adds a trap potential $\sum_{i=1}^N
V(\x_i)$ to the Hamiltonian (\ref{ham}). A typical example, which
describes the experimental situation rather well, is a harmonic
oscillator potential $V(\x) = \omega^2 |\x|^2$, with $\omega>0$ the
trap frequency. More generally, the trapping frequencies in the three
directions can be different, of course.

A characteristic feature of these trapped gases is their response to
rotation. One observes the appearance of quantized vortices
\cite{madison,engels}, whose number increases with the rotation speed
$|\vec \Omega|$.

Even a rotating Bose gas can be described in a time-independent way,
by going to the rotating reference frame. The only effect on the
Hamiltonian is to add the term $\sum_{i=1}^N \vec\Omega\cdot \vec
L_i$, as discussed in the Introduction. To ensure stability of the system, the trap potential $V$ has to increase fast enough at infinity to compensate for the centrifugal force in the rotating system. More precisely, we have to assume that 
$$
\lim_{|\x|\to \infty} \left( V(\x) - \tfrac 14 |\vec\Omega\wedge \x|^2\right) = +\infty\,.
$$

\subsection{The Gross-Pitaevskii Equation}

The previous considerations suggest that dilute Bose gases close to zero temperature should be well described by the Gross-Pitaevskii (GP) energy functional \cite{Gross,Pitaevskii}\begin{equation}\label{gpf}
{\mathcal E}^{\rm GP}[\phi] = \left\langle \phi\left| -\Delta + V(\x) - \vec\Omega\cdot \vec L  \right|\phi\right\rangle + 4\pi Na \int_{\R^3} |\phi(\x)|^4 d\x \, .
\end{equation}
Its ground state energy is  
$$
E^{\rm GP}(Na,\vec\Omega)= \inf_{\|\phi\|_2=1} {\mathcal E}^{\rm GP}[\phi]\,,
$$
and any minimizer satisfies the GP equation
$$
\boxed{ \phantom{\int} 
-\Delta\phi(\x) +V(\x)\phi(\x) - \vec\Omega\cdot \vec L \, \phi(\x)  + 8\pi Na |\phi(\x)|^2 \phi(\x) = 
\mu \phi(\x)\,.\quad\ } 
$$
For a minimizer, $N|\phi(\x)|^2$ is interpreted as the particle
density of the system. Hence the last term in (\ref{gpf}) is the
natural generalization of the expression $4\pi a \rho^2$ to
inhomogeneous systems.

For $\vec\Omega\neq 0$ and axially symmetric $V(\x)$, the rotation
symmetry can be broken due to the appearance of quantized
vortices. More precisely, it was shown in \cite{S02,S03} that for all trap
potentials $V(\x)$ that grow faster than quadratically at infinity,
there exists a $g_{\vec \Omega}$ such that for all $Na > g_{\vec
  \Omega}$ the GP minimizers necessarily are not axially symmetric. In
particular, there are many (in fact, uncountably many) GP minimizers!
The symmetry breaking is due to the appearance of quantized vortices which can not
be arranged in a symmetric way. Many interesting results have been
obtained concerning the nature and distribution of these vortices in
GP minimizers. We refer to \cite{aft} and references therein.

\subsection{Ground State Energy of Dilute Trapped Gases}

In typical experiments on cold atomic gases, $N\gg 1$, $a\ll 1$ (the length scale of the trapping potential $V$), but $Na=O(1)$. To get to this dilute regime, one writes 
\begin{equation}\label{scv}
v(\x) = \frac 1 {a^2} w(\x/a)
\end{equation}
with $w$ having scattering length $1$. It is easy to see that $v(\x)$ then has scattering length $a$. The scattering length thus becomes a parameter, and we can write
$$
\infspec H = E_0(N,a,\vec\Omega)\,.
$$
We note that the scaling (\ref{scv}) is of course equivalent to a
rescaling of the trap potential $V$ while keeping $v$ fixed. This
latter procedure may seem physically more natural (as the trap
potential is easier to adjust experimentally than the interaction
potential) but we find it more convenient to fix $V$ instead and scale
$v$ as in (\ref{scv}) instead. Our procedure corresponds to measuring
all lengths in the system in units of the length scale of the trap
potential.

For dilute systems, one expects that $E_0(N,a,\vec\Omega) \approx N E^{\rm GP}(Na,\vec\Omega)$. The proof of this fact was given in \cite{LiebS06}.

\begin{theorem}[\str{Ground State Energy of Trapped Gases}]
For fixed $g\geq 0$ and $\vec\Omega\in \R^3$, 
\begin{equation}\label{gpr}
\boxed{\ \lim_{N\to\infty} \frac {E_0(N,g/N,\vec\Omega)}{N} = E^{\rm
    GP}(g,\vec\Omega) \ }
\end{equation}
\end{theorem}

This theorem was previously proved in \cite{LSY00} for the case
$\vec\Omega=0$. The main difficulty in the generalization to rotating
systems comes from the fact that the permutation symmetry of the wave
functions now becomes essential. While for non-rotating systems it is
well known that the ground state for bosons coincides with the ground
state without symmetry restrictions (as the latter is unique and positive,
hence must be symmetric), this fact fails to hold for rotating
systems. In fact one can show that (\ref{gpr}) fails to hold, in
general, if the left side is replaced by the absolute ground state
energy of $H$ (viewed as an operator on $L^2(\R^{3N})$, without
symmetry restrictions).\cite{S03}

\subsection{BEC for Rotating Trapped Gases}

In the previous subsection it was argued that the ground state
energy of the GP functional (\ref{gpf}) is a good approximation to the
ground state energy of $H$ for dilute gases.  For the corresponding
ground state $\Psi_0(\x_1,\dots,\x_N)$, one would also expect that its one-particle reduced density matrix satisfies
\begin{equation}\label{becrot}
\gamma_0 \equiv  \Tr^{(N-1)} |\Psi_0\rangle\langle \Psi_0| \approx |\phi\rangle\langle \phi|
\end{equation}
with $\phi$ a minimizer of the GP functional. (For convenience, the
normalization of $\gamma_0$ has been chosen differently here than we
did previously in (\ref{def:rdm}).) While this is indeed true in the
non-rotating case \cite{LS02}, the rotating case is more complicated
because of the non-uniqueness of the GP minimizers $\phi$. The best
one can hope for is to replace the right side of (\ref{becrot}) by a
convex combination of rank-one projections onto GP minimizers. This is
indeed the content of Theorem~\ref{condensation} below, which was
proved in \cite{LiebS06}.

To state the following results precisely, it is necessary to introduce
the concept of an approximate ground state. We will call a sequence of
$N$-particle density matrices (positive trace class operators on the
$N$-particle space with trace equal to one) an approximate ground
state if their energy equals the ground state energy to leading order
in $N$. Then we define the set $\Gamma$ as the set of limit points of
one-particle density matrices of such approximate ground states. More
precisely,
\begin{equation}\label{gamma}
\Gamma= \left\{ \gamma\, : \, \exists {\rm \ sequence\ } \gamma_{N},
\lim_{N\to\infty, \, Na\to g} \frac 1N \Tr\, H \gamma_{N}= E^{\rm GP}(g,\vec\Omega),\
\lim_{N\to\infty} \gamma_{N}^{(1)}=\gamma\right\}
\end{equation}
where $\gamma_N^{(1)} = \Tr^{(N-1)} \gamma_N$ denotes the one-particle density matrix of $\gamma_N$.

\begin{theorem}[\str{BEC for Dilute Trapped Gases}]  \label{condensation}
The set $\Gamma$ in (\ref{gamma}) has the following properties.
\begin{itemize}
\item[(i)] $\Gamma\subset \mathcal{J}_1 $ is compact and convex. 

\item[(ii)] The extreme points $\Gamma_{\rm ext}\subset \Gamma$ are given by GP minimizers, i.e.,
 $\Gamma_{\rm
    ext} = \{ |\phi\rangle\langle\phi|\, : \, {\mathcal E}^{\rm GP}[\phi]=E^{\rm
    GP}(g,\vec\Omega)\}$.
\item[(iii)] For every $\gamma\in \Gamma$ there exists a positive (regular Borel) measure  $d\mu_\gamma$, supported in $\Gamma_{\rm ext}$ with $\int_{\Gamma_{\rm ext}} d\mu_\gamma(\phi) =1$, such that  
\begin{equation}\label{gdm}
\gamma = \int_{\Gamma_{\rm ext}} d\mu_\gamma(\phi)\,
|\phi\rangle\langle\phi|\,.
\end{equation}
\end{itemize}
\end{theorem}

Eq.~(\ref{gdm}) is the natural generalization of (\ref{becrot}) to the
case of multiple GP minimizers. It says that the one-particle density
matrix of any approximate ground state is close (in trace class norm)
to the convex combination of projections onto GP minimizers.

Theorem~\ref{condensation} represents also a proof of the spontaneous
breaking of the rotation symmetry in rotating Bose gases. An
infinitesimal perturbation, e.g. of the trap potential $V$, leads to a
unique GP minimizer and hence to 100\% condensation, since the set
$\Gamma$ consists of only one element in this case. The quantized
vortices are visible in the GP minimizer; they are a typical feature
of superfluids. Theorem~\ref{condensation} can therefore also be
interpreted as a proof of the superfluid behavior of rotating Bose
gases.\cite{LSYsuperflu}

\subsection{Rapid Rotation}
Consider now the special case of a harmonic trapping potential
$$
V(\x) =  \tfrac 14  |\x|^2\,.
$$
As discussed above, $H$ is bounded below only for $|\vec\Omega|\leq
1$. The results in the previous subsections are valid for fixed
$|\vec\Omega|<1$. The question we would like to address in this final
section is what happens as $|\vec\Omega|\to 1$? Denoting $\vec
e_\Omega=\vec\Omega/|\vec\Omega|$ the unit vector in the direction of
$\vec\Omega$, we can write
\begin{equation}\label{ope}
-\Delta + \tfrac 14 |\x|^2 - \vec\Omega\cdot \vec L = \underbrace{ \left(-i\vec\nabla -
  \half \vec e_\Omega\wedge \x\right)^2 + \tfrac 14 |\vec e_\Omega\cdot \x|^2 }_h +
(\vec e_\Omega-\vec\Omega)\cdot \vec L\,.
\end{equation}
The operator $h$ has eigenvalues $\tfrac 32, \tfrac 52, \tfrac 72,
\dots$, each of which is infinitely degenerate.

For low energies it makes sense to restrict the allowed wave functions to the kernel of $h - \tfrac 32$. This kernel is given by the Bargmann space \cite{bargmann}
\begin{equation}\label{bs}
\{ f(z) e^{-|\x|^2/4} \, , \ f:\C\to \C \ \text{analytic} \} \subset L^2(\R^3) 
\end{equation}
where we identify the complex variable $z$ with the plane perpendicular to $\vec\Omega$. In particular, $|\x|^2 = |z|^2 + |\vec e_\Omega\cdot \x|^2$. Since the Gaussian factor is fixed, it is convenient to absorb it into the measure and think of the Hilbert space as a space of analytic functions only. It can easily be checked that the angular momentum operator $\vec e_\Omega\cdot \vec L$ acts on $f$ as $z\partial_z$. In particular, its eigenfunctions are $z^n$, with eigenvalues $n \in \{0,1,2,\dots\}$.

We note that if one interprets $\vec e_\Omega$ as a homogeneous magnetic field, the Bargmann space (\ref{bs})  corresponds to the lowest Landau level in the perpendicular direction, multiplied by a fixed Gaussian in the longitudinal direction.

Now that we have identified the Bargmann space (\ref{bs}) as the
appropriate one-particle Hilbert space for rapidly rotating bosons, we
have to come up with an effective Hamiltonian describing this
system. The only term left in the one-particle energy (\ref{ope}),
expect for a trivial factor $\tfrac 32$, is the angular momentum term
$(\vec e_\Omega -\vec \Omega)\cdot \vec L$. If the range of the
interaction potential is much shorter than the \lq\lq magnetic
length\rq\rq\ $1$, it makes sense to approximate the interaction
potential by a $\delta$-function, which becomes a bounded operator
when projected to the Bargmann space. Writing the prefactor of the
$\delta$-function as $8\pi a$, in accordance with previous
considerations, we arrive at the effective Hamiltonian
\begin{equation}\label{hlll}
\boxed{ \ H^{\rm LLL} : = (1-|\vec\Omega|) \sum_{i=1}^N z_i \partial_{z_i} + 8\pi a
\sum_{1\leq i<j\leq N} \delta_{ij} \,. \ } 
\end{equation}
It acts on the space of permutation-symmetric analytic functions
$f(z_1,\dots,z_N)$ which are square-integrable with respect to the
measure $\prod_{i=1}^N e^{-|z_i|^2/4} dz_i$. We denote this space by
$\mathcal{B}^{\otimes N}$. The operator
$\delta_{12}:\mathcal{B}^{\otimes 2}\to \mathcal{B}^{\otimes 2}$ acts
as
$$
\left( \delta_{12} f\right)(z_1,z_2) = \frac 1{(2\pi)^{3/2}} f\left( \tfrac 12 (z_1+z_2),\tfrac 12 (z_1+z_2) \right)
$$
which takes analytic functions into analytic functions.  It is
obtained by projecting $\delta(\x_1-\x_2)$ onto $\mathcal{B}^{\otimes
  2}$.

Concerning the effective Hamiltonian (\ref{hlll}), the following questions arise naturally:

\begin{itemize}
\item [(1)] Can one derive $H^{\rm LLL}$ rigorously from the full
  Hamiltonian $H$ as $|\vec\Omega|\to 1$ and $a\to 0$? Such a rigorous
  derivation was indeed achieved in \cite{LewinS2009}, where it was
  shown that if $|\vec \Omega|\to 1$ for fixed $N$ and fixed
  $a/(1-|\vec\Omega|)$ (i.e., also $a\to 0$), then the ratio of the
  ground state energy of $H$, minus the trivial term $\tfrac 32 N$, to
  the ground state energy of $H^{\rm LLL}$ goes to 1. Similarly, one
  can show that also eigenfunctions converge in the same
  limit. Uniformity in the particle number $N$ is still an open
  problem, however.

\item [(2)] What are the properties of $H^{\rm LLL}$, in particular concerning its spectrum and corresponding eigenfunctions? Certain features of $H^{\rm LLL}$ are expected to show some similarities to the fractional quantum Hall effect which occurs in fermionic systems.
\end{itemize}

Concerning the latter question, let us first note that the two terms  $\mathcal{L}_N$ and $\Delta_N$ in $H^{\rm LLL}$ commute:
$$
H^{\rm LLL} = (1-|\vec\Omega|) \underbrace{ \sum_{i=1}^N z_i \partial_{z_i}}_{\mathcal{L}_N} + 8\pi a
\underbrace{\sum_{1\leq i<j\leq N} \delta_{ij} }_{\Delta_N}\,.
$$
Hence the ground state energy $E_0^{\rm LLL}(N,a,\vec\Omega) = \infspec H^{\rm LLL}$ is obtained from the joint spectrum of these two operators. Of particular relevance is the  yrast curve, which is defined as the lowest eigenvalue of $\Delta_N$ in the sector of total angular momentum $L$:
$$
\Delta_N(L) = \infspec \Delta_N \restriction_{\mathcal{L}_N=L} \,.
$$
It is explicitly known for $L\leq N$ \cite{bp,pb,hv} (see also \cite{LewinS2009} for a simple proof)
$$
\Delta_N(L) = (2\pi)^{-3/2} \left\{ \begin{array}{ll} \tfrac 12 N(N-1) & \text{for $L\in\{0,1\}$} \\ \tfrac 12 N \left(N-1-\tfrac 12 L\right) & \text{for $2\leq L \leq N$.} \end{array}  \right. 
$$
Moreover, $\Delta_N(L)=0$ for $L\geq N(N-1)$. The eigenfunctions for $L=N(N-1)$ corresponding to the eigenvalue $0$ of $\Delta_N$ is the bosonic Laughlin wave function
$$
\prod_{1\leq i<j\leq N} (z_i-z_j)^2\,.
$$

\begin{figure}[ht]
\small
\begin{center}
\input{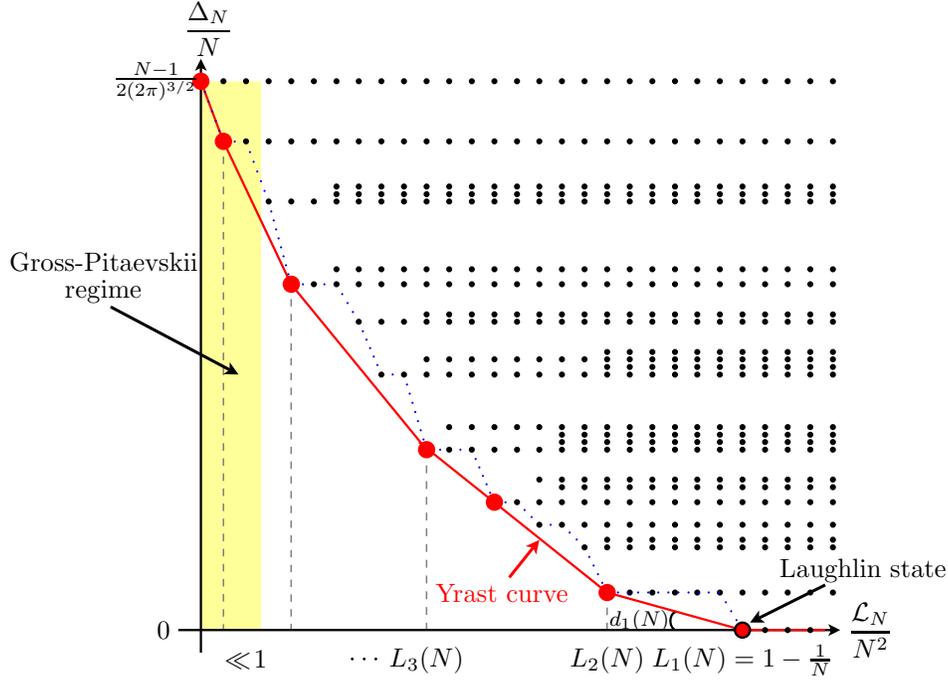}
\end{center}
\caption{(taken from \cite{LewinS2009}) A sketch of the joint spectrum of $\mathcal{L}_N$ and $\Delta_N$. The dotted line is the yrast curve, its convex hull is in red. The bold dots correspond to the possible ground states of $H^{\rm LLL}$ as one varies $(1-|\vec\Omega)/a$.  The yellow area on the left shows the validity regime of the Gross-Pitaevskii equation. For $L\geq N(N-1)$, the interaction energy is zero.}\label{fig:yr}
\end{figure}

Little is known about $\Delta_N(L)$ for $N< L <N(N-1)$, except for numerical simulations for small particle number. The only rigorous result concerns the limit $N\gg 1$ and $L\ll N^2$ where one can show that the Gross-Pitaevskii approximation is exact \cite{LiebSYngvason2009}. I.e., in this regime the convex hull of  $\Delta_N(L)$ is given by 
$$
\inf\left\{ \tfrac 12  \left\langle f\otimes f\left| \delta_{12} \right| f\otimes
    f\right\rangle   \, : \, f \in \mathcal B,\, \|f\|^2=N ,\, \left\langle f\left| z\partial_z \right| f\right\rangle = L \right\}\,.
$$
The qualitative behavior of $\Delta_N(L)$ is sketched in Figure~\ref{fig:yr}.

\section*{Acknowledgments}

This article was written during a visit at the Erwin-Schr\"odinger Institute in Vienna, whose hospitality is gratefully acknowledged. This work was partially supported by the U.S. National Science Foundation under grant No. PHY-0652356.

\end{document}